\documentclass[%
 reprint,
 amsmath,amssymb,
prb,
]{revtex4-1}

\usepackage{graphicx}
\usepackage{dcolumn}
\usepackage{bm}

\begin{document}
\newcommand{\Tr}{$T_\text{1}$}
\newcommand{\Ts}{$T_\text{2}$}
\newcommand{\Qr}{$Q_\text{R}$}
\newcommand{\Qru}{$Q_\text{Ru}$}
\newcommand{\Qs}{$Q_\text{S}$}
\newcommand{\Vm}{$V_\text{M}$}
\newcommand{\Vb}{$V_\text{B}$}
\newcommand{\Vg}{$V_\text{G}$}
\newcommand{\Vgm}{$V_\text{G}^\text{M}$}
\newcommand{\Ic}{$I_\text{C}$}
\newcommand{\Ib}{$I_\text{B}$}
\newcommand{\fd}{$f_\text{d}$}
\newcommand{\fp}{$f_\text{p}$}
\newcommand{\degree}{\ensuremath{^\circ}}

\title{Observation of decoherence in a carbon nanotube mechanical resonator}

\author{Ben H. Schneider}
\email{benh.schneider@gmail.com}
\author{Vibhor Singh}
\author{Warner J. Venstra}
\author{Harold B. Meerwaldt}
\author{Gary A. Steele}
\email{g.a.steele@tudelft.nl}
\affiliation{Kavli Institute of Nanoscience, Delft University of Technology, 2628 CJ Delft, The Netherlands}

\date{\today}

\begin{abstract}
In physical systems, decoherence can arise from both dissipative and dephasing processes.
In mechanical resonators, the driven frequency response measures a combination of both,
while time domain techniques such as ringdown measurements can separate the two.
Here, we report the first observation of the mechanical ringdown of a carbon nanotube mechanical resonator.
Comparing the mechanical quality factor obtained from frequency- and time-domain measurements, we find a spectral quality factor four times smaller than that
measured in ringdown, demonstrating dephasing-induced decoherence of the nanomechanical
motion. This decoherence is seen to arise at high driving amplitudes,
pointing to a non-linear dephasing mechanism. Our results highlight
the importance of time-domain techniques for understanding dissipation
in nano-mechanical resonators, and the relevance of decoherence mechanisms in
nanotube mechanics.
\end{abstract}

\maketitle

\vspace*{\fill}
\clearpage
Decoherence in mechanical resonators corresponds to a loss of phase information of the oscillations in position, similar to decoherence in quantum systems due to the loss of phase information of a superposition state. In contrast, dissipation in a mechanical oscillator or quantum system corresponds to a loss of energy over time\cite{cleland2002noise, marquardt2008introduction, remus2009damping}.
The distinction between decoherence and dissipation is often overlooked when studying the quality factor of a mechanical resonator. The mechanical response is typically characterised by a quality factor $Q$, which is measured from either a spectral response or by a ringdown measurement. In a spectral response measurement, the driving frequency of the resonator is swept and the steady-state oscillation amplitude is measured. From the width of the response peak, a quality factor is extracted. In a ringdown measurement, a quality factor is determined by the envelope of the transient of the resonator after a driving force is switched off.

Despite the fact that they are both given the same symbol $Q$, the
quality factor as obtained from these two types of measurements are
sensitive to different physical processes. 
The quality factor measured in the spectral response, which we will denote here as \Qs, is sensitive to
both dissipation (energy loss) and pure dephasing (such as fluctuations of the resonators resonance frequency), similar to the \Ts~time in a Ramsey
experiment with a qubit\cite{Ramsey1950}. 
The quality factor \Qr~measured in a
ringdown experiment is sensitive only to dissipation (energy loss), similar to a \Tr~measurement on a qubit\cite{Bloch1946}. In qubits, it is well-known that the \Tr~and \Ts~can be very different: an extreme example is a GaAs spin
qubit, in which \Ts~can be as short as 10~ns, while \Tr~can be as long as 1~s\cite{petta2005coherent,amasha2008electrical}.  

Decoherence in mechanical resonators has been studied earlier in a piezoelectric resonator coupled to a superconducting circuit\cite{o2010quantum}, in atomically thin drum resonators\cite{van2014time}, and in the two coupled modes of a vibrating string\cite{faust2013coherent}. 
In these experiments, no pure dephasing was observed and the coherence of the motion was limited by dissipation. In experiments with mass sensing\cite{yang2011surface}, excess phase noise was observed that which was attributed to surface diffusion of molecules\cite{Atalaya2011}. 
In comparison to conventional NEMS, carbon nanotube resonators are very sensitive to their environment. 
Due to a low mass and spring constant, 
nanotube mechanical resonators show a strong dispersion with the gate voltage\cite{sazonova2004tunable} are sensitive to the force of a single electron charge\cite{lassagne2009coupling,steele2009strong}, exhibit strong mode coupling\cite{eichler2012strong,castellanos2012strong}, and strong nonlinearities\cite{eichler2011nonlinear}. 
These strong coupling effects may give rise to dephasing, making them an interesting candidate for exploring mechanical decoherence.

\begin{figure}[ht]
\includegraphics{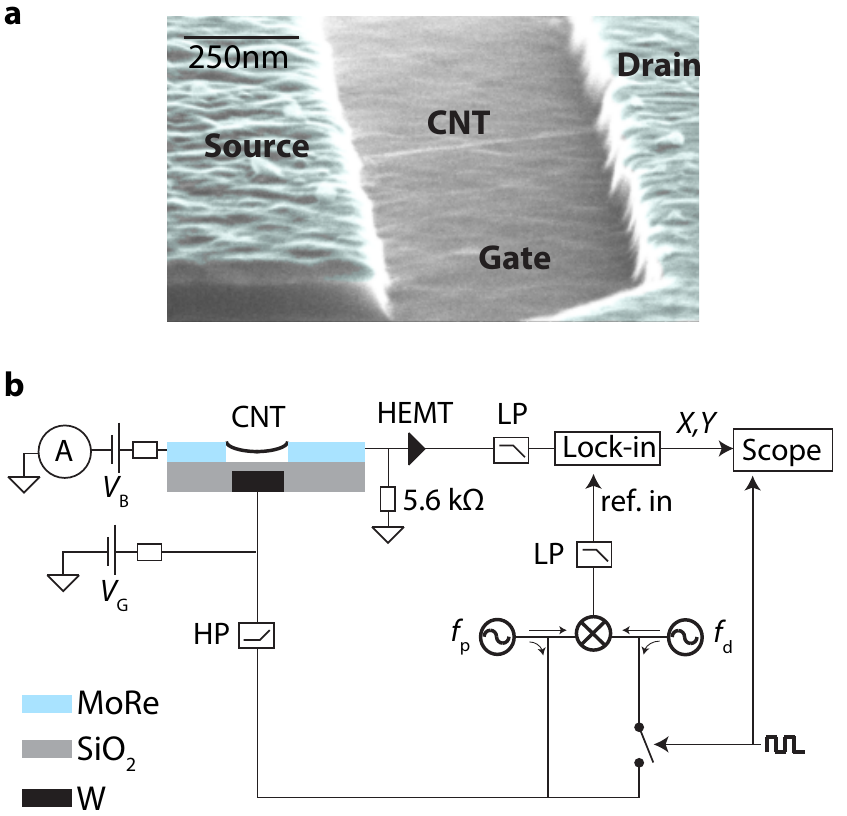}
\caption{\label{rd:fig1}
\textbf{Device and measurement setup}
(\textbf{a}) Colorised scanning electron microscope image of a typical
device, showing a suspended CNT in contact with source and drain
electrodes. (\textbf{b}) Schematics of the measurement setup. To measure the mechanical response, two
frequency signal generators are used: one for the drive and one for
the probe signal.  The probe signal \fp~is detuned by 7~MHz from the drive
signal \fd.  A RF radio frequency switch is used to turn on and off the drive signal to the
gate using square-wave pulses. A mixing signal is generated with a mixer at room temperature
which serves as a reference signal for the lock-in amplifier.  A second mixing signal is generated by the CNT, which is impedance-matched by a High Electron Mobility Transistor (HEMT) and detected by a lock-in amplifier.  The $X$ and $Y$ quadrature outputs from the lock-in are recorded in sync with the switch by an oscilloscope. In the schematic, HP (LP) represent high (low) pass filters.}
\end{figure}

Furthermore, the low value of the quality factor in carbon nanotube resonators is not well understood.
It has been proposed that low $Q$ factors could originate from thermally induced spectral broadening\cite{barnard2012fluctuation}, clamping losses\cite{cole2011phonon, rieger2014energy,aykol2014clamping}, or from symmetry breaking\cite{eichler2013symmetry}.  
Until now, reports of the quality factor in carbon nanotubes have been
based only on spectral measurements \cite{huettel2009carbon,lassagne2009coupling,eichler2011nonlinear}.
To distinguish dissipation processes from dephasing, additional measurements, such as a mechanical ringdown, must be performed.

Here, we use a recently developed fast detection scheme to measure the ringdown of a carbon nanotube mechanical resonator\cite{meerwaldt2013submicrosecond}.  
Doing so, we gain access to both the dissipation and decoherence processes.
At low driving power, we find that \Qs~and \Qr~have the same value,
indicating that dephasing processes are not significant.  
At higher driving power however, we find that spectral response becomes
significantly broader, with a drop in \Qs~by a factor of four while \Qr~remains constant, demonstrating decoherence from dephasing of the nanotube motion.\vspace{10pt}
\clearpage

{\bf \Large Results}
\\

\begin{figure}[ht]
\includegraphics{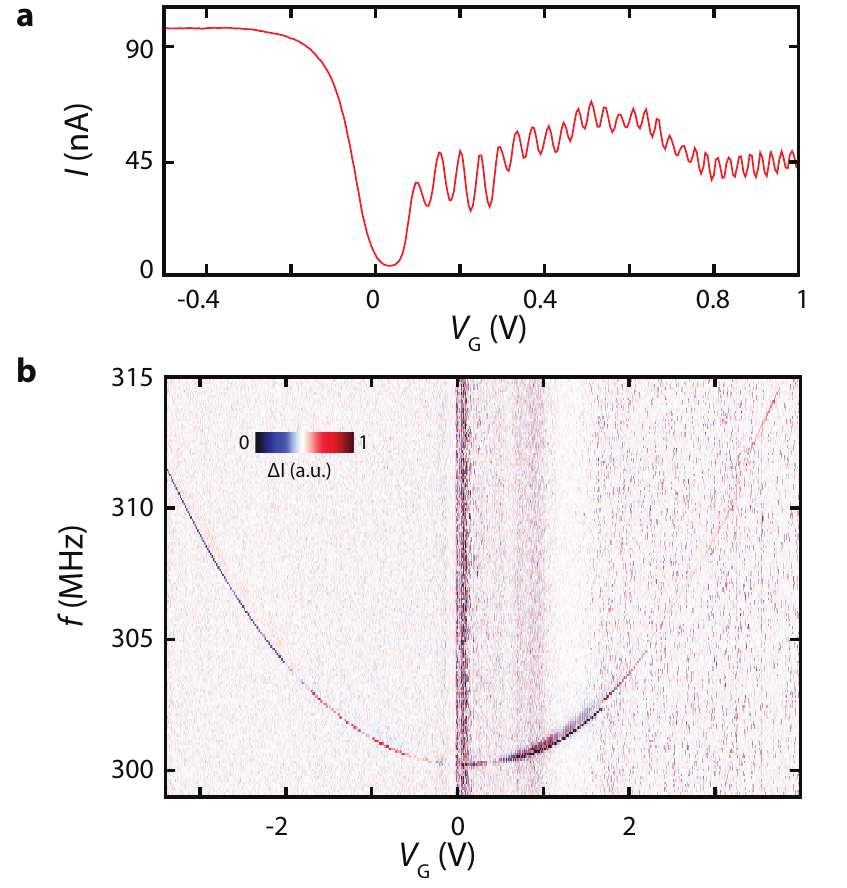}
\caption{\label{rd:fig2} 
\textbf{Device characterisation.}
(\textbf{a}) Conductance as a function of gate voltage \Vg at 2 Kelvin and at \Vb $= 4$~mV. 
For \Vg~$< 0$~V the conductance is high while for \Vg~$>0$~V the conductance is low, exhibiting weak Coulomb blockade.
(\textbf{b}) Resonance frequency as a function of gate voltage,
showing that the suspended CNT acts as a gate-tuneable mechanical
resonator.}
\end{figure}

\noindent {\bf Description and characterization of the device} \\
\noindent  A suspended single-walled carbon nanotube (CNT) mechanical resonator
is fabricated as described previously\cite{meerwaldt2013submicrosecond}.  
Briefly (Supplementary Methods), fabrication starts with an intrinsic silicon wafer, on which a gate
electrode is patterned.  On top, a 200~nm thick silicon oxide layer is
deposited, followed by 70~nm thick MoRe source and drain electrodes.
Finally, a CNT is grown across the 500~nm trench that separates the
electrodes.  The distance between the gate and the CNT is 270~nm.
Figure \ref{rd:fig1}(a) shows a scanning electron microscope (SEM)
image of a typical device.
Figure \ref{rd:fig1}b shows the schematic of the circuit used to measure the conductance of the CNT as a function of gate voltage (Fig.~\ref{rd:fig2}a). 
The measurements presented here are performed at 2~K.
A high conductance is observed for negative gate voltages, when the CNT is doped with holes.  
When a positive gate voltage is applied, a low conductance region (small band
gap) is followed by weak conductance oscillations due to Coulomb
blockade.  This overall behaviour is typical for a clean single CNT in
contact with MoRe electrodes\cite{schneider2012coupling}.

To find the mechanical resonance and to characterise its gate
dependence, we first measure the nanotube motion using the
rectification technique outlined in Ref.\onlinecite{steele2009strong}.
Figure \ref{rd:fig2}b shows the resonant change in DC current $\Delta I$ as a function of
frequency \fd~and \Vg~at \Vb = 1~mV.  
The frequency at which the mechanical signal is detected increases with gate voltage, characteristic for a carbon nanotube mechanical resonator\cite{sazonova2004tunable}. 
The approximately quadratic gate voltage dependence suggests that there is some slack in the nanotube.\\

\begin{figure}[ht]
\includegraphics{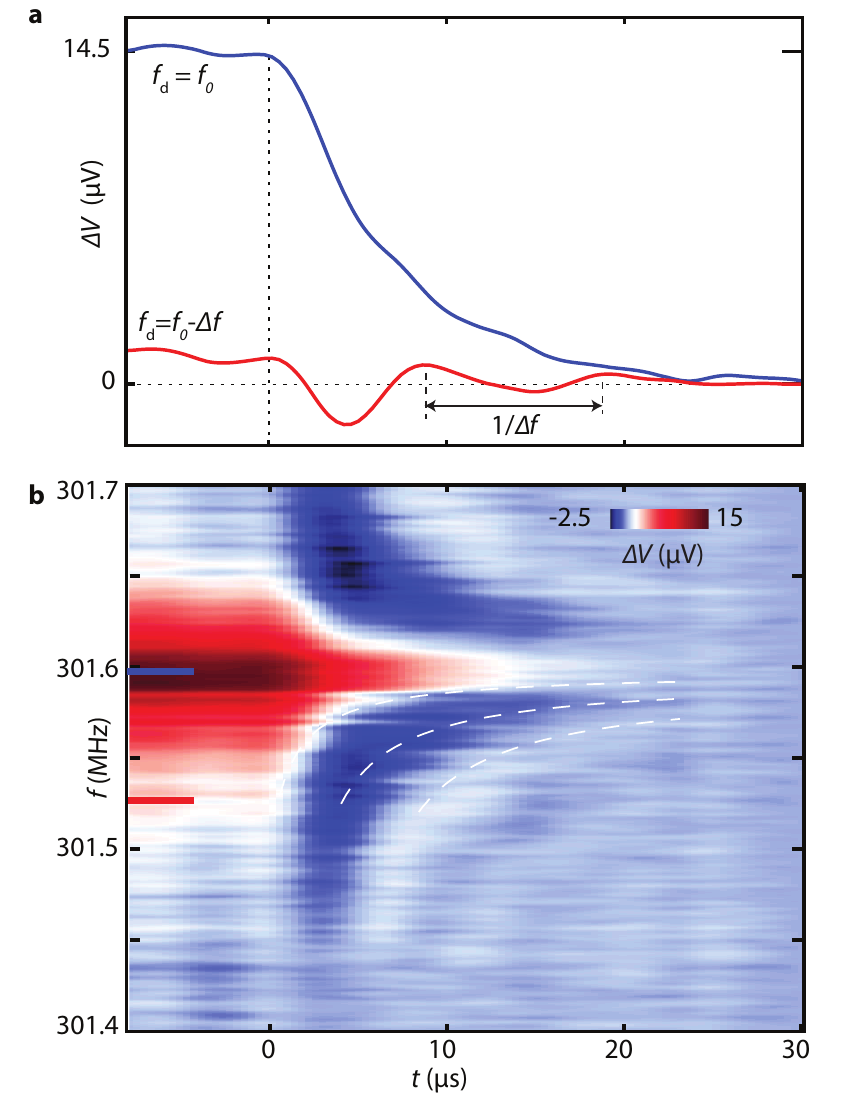}
\caption{\label{rd:heidoscil}
\textbf{Time domain response of the CNT motion.}
(\textbf{a})
Recorded response of the lock-in amplifier when switching off the
drive signal for an on resonance (blue) and off-resonance (red) drive frequency.
The red curve shows the mechanical response of the CNT when the drive signal is detuned by $\Delta f = 70$~kHz from the mechanical resonance position. An oscillation on top of the
ringdown signal (for $t > 0$) is visible.  The oscillation periodicity matches with $P =
\frac{1}{\Delta f}$.  
(\textbf{b}) Colour-scale map of the
mechanical ringdown signal as a function of time and drive frequency.
A white contour on top of the colour-scale map indicates the peak and dip
positions due to the detuning from the fundamental mechanical
resonance frequency.  The red and blue markers in (\textbf{b})
the positions where the line-cuts in \textbf{a} are extracted from.}
\end{figure}

\noindent {\bf Observation of the mechanical ringdown of a carbon nanotube}\\
\noindent In figure \ref{rd:heidoscil}, we measure the response of the mechanical resonator in the time domain.
The schematic of the detection circuit is shown in Fig.~\ref{rd:fig1}b.  
Two signal generators are used to generate a drive (\fd) and a probe (\fp)
signal. The amplitude of the drive signal was chosen to be sufficiently small such that mechanical response does not exhibit a nonlinear Duffing line-shape. The probe signal is 7~MHz detuned from the drive frequency
and a mixer is used to generate a 7~MHz reference signal for the
lock-in.  The CNT resonator is driven and detected by applying the drive and probe
signals to the gate.
The transconductance of the CNT leads to the mixing of these two signals (see Supplementary Note 1, 2 and Supplementary Fig. 1-3).
Note that this is slightly different than conventional two-source mixing where one
signal is applied to the source and the other to the gate:
at cryogenic temperatures the nonlinear response of the nanotube conductance with gate voltage $G$(\Vg) allows us to use a similar mixing-type detection \cite{sazonova2004tunable} with signals applied only to the gate and a constant DC voltage bias \Vb = -5~mV.
To impedance-match the signal coming from the nanotube, 
a HEMT is located in close proximity to the CNT, 
so that the motion can be detected with a bandwidth of 62~MHz\cite{meerwaldt2013submicrosecond}. Here we use a high-frequency lock-in amplifier to detect the signal from the HEMT amplifier, in which case our readout scheme can detect motion on $\mu$s timescales.

To measure the mechanical ringdown, we use the following pulsed
excitation and measurement scheme.  A switch in the circuit controls
the drive signal (\fd) which is connected to the gate of device.  
With the switch on, the motion of the CNT is excited.
When the switch is turned off, only the probe and an effective gate voltage from the motion of the carbon nanotube are present.
These two are mixed by the CNT, and the resulting signal is amplified by the HEMT and detected by the lock-in amplifier.
The envelope of the mechanical ringdown signal is then detected as a function of time by 
detecting the output of the lock-in amplifier with an oscilloscope.

The blue curve in figure \ref{rd:heidoscil}a shows the transient response measured by the lock-in after the CNT resonator has been driven at resonance (\fd = $f_0$).  
The gate voltage is fixed at \Vg = 0.4~V and the drive signal is switched off at $t = 0~\mu$s. 
To improve the signal-to-noise ratio, the output quadratures from the lock-in have been averaged using the oscilloscope, typically 10,000 times.
The averaging was done by applying 
a repeating square-wave voltage signal to the radio-frequency switch Fig.~\ref{rd:fig1}.
The same square-wave is used to trigger the oscilloscope.
In the data shown, a constant offset from electrical mixing has been subtracted, and the quadratures have been rotated (Supplementary Note 2 and Supplementary Figs. 4-7) such that the data presented represent the amplitude of the mechanical signal. The rounding of the decay curve around $t=0$ is consistent with smoothing on a short timescale from the lock-in filter response (See Supplementary Note 3 and Supplementary Fig. 8).

The red curve in Fig.~\ref{rd:heidoscil}a shows the response when the drive signal is detuned by $\Delta f \sim70~$kHz from the mechanical resonance of the CNT\cite{antoni2012nonlinear}.  
In this case the signal does not just decay but instead oscillates below the
relative zero position, while decreasing in amplitude.  This
oscillation frequency matches the frequency detuning relative to the
mechanical resonance frequency ($\Delta f = f_0 - f$). 
The origin of these oscillations lies in the behaviour of a driven mechanical resonator.
When driven off-resonance, the motion oscillates at the driving frequency.
However, when the driving is switched off, the mechanical resonator oscillates at its natural frequency.
Because the reference of the lock-in amplifier is referenced by the initial drive frequency, the output of the lock-in oscillates at the difference frequency $\Delta f$ (see Supplementary Note 2 and Supplementary Fig. 7).
The observation of this oscillation is a clear confirmation that the observed transients are mechanical in nature.

To further investigate this behaviour, we plot in figure
\ref{rd:heidoscil}b a two-dimensional colour-scale representation of
the measured amplitude as a function of time and drive frequency.
While keeping the gate voltage \Vg = 0.4~V fixed, we slowly step the
drive and probe signal across the mechanical resonance of the CNT.
For each frequency we record the averaged lock-in output
quadratures with the oscilloscope.  On top of the colour-scale map, a
contour map with white dashed lines is shown, 
to indicate the positions of the peaks and dips of the
ringdown signals that are recorded off-resonance.  
The, dashed lines indicate the expected peak and dip positions
with a periodicity of $\frac{1}{\Delta f}$, where $\Delta f =
f_0 - f_d$ is the detuning relative from the mechanical resonance position. 
As can be seen from the 2D colour map, 
the dashed lines follow the observed peak/dip positions, 
confirming that the oscillating ringdown signal is indeed from the mechanical resonator.\\

\begin{figure}[ht]
\includegraphics{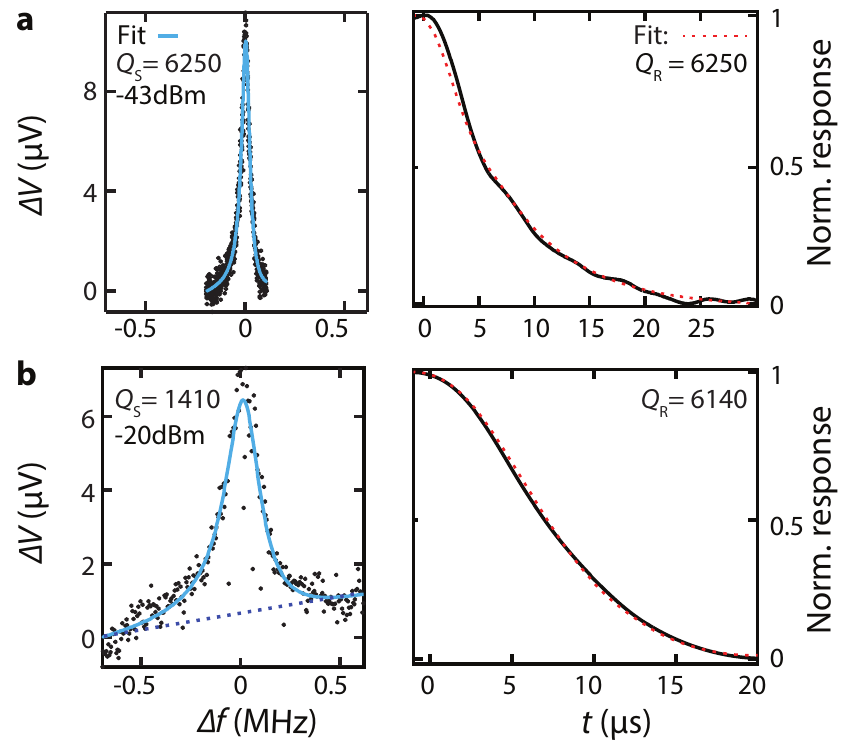}
\caption{\label{rd:powdep} 
\textbf{Decoherence of a carbon nanotube mechanical resonator.}
The gate voltage \Vg = 0.4~V and the temperature is 1.9~K. 
The drive power and fitted quality factor are indicated in the panels.
In (\textbf{a}) the probe power is -28~dBm and the resonance position $f_0 = 301.6$~MHz.  
In (\textbf{b}) the probe power is -8~dBm and the resonance position $f_0 = 299.69$~MHz.  
The data was fitted by including a slowly varying frequency dependent electrical mixing signal (dashed line), see Supplementary Note 3 and Supplementary Fig. 9, 10. We estimate an amplitude of 0.14~nm and 0.7~nm for \textbf{a} and \textbf{b}, respectively (Supplementary Note 2).}
\end{figure}

\noindent {\bf Decoherence and non-linear dephasing of the motion of a carbon nanotube}\\
\noindent To extract the ringdown quality factor \Qr, we fit the measured time response of device for resonant driving. An important detail in the experiment is that the averaging in the oscilloscope is performed on time traces of the amplitude quadrature. To describe such quadrature-averaged measurements, we introduce a model that includes the effects of dephasing and of the filter response of the lock-in amplifier (see Supplementary Note 3 and Supplemary Figs 8-10). 
This model, which is constrained by the independently observed \Qs, 
allows us to extract \Qr~from the mechanical transient.
In Fig. \ref{rd:powdep}, we use this technique to independently extract the dephasing and dissipation contributions to the mechanical quality factor. The upper panels show the spectral response and the time domain response for a low driving power. Fitting the datasets, we find that both \Qs~and \Qr~are well-described by a single number (\Qs = \Qr~$\sim 6250$) indicating that dephasing does not play a role at this driving power.

In the lower panels, we show the mechanical response at higher driving powers. At higher driving powers, the spectral response is still well-described by the line-shape of an harmonic oscillator (see
Supplementary Note 1 and Supplementary Fig. 1) with no sign of hysteresis, indicating that the amplitude of the motion is small enough that nonlinear restoring forces (which would lead to a Duffing response) are not significant. Although the response still fits well to the line-shape of an harmonic oscillator, it has increased in line-width, now exhibiting a \Qs~$\sim 1410$. In the right panels, we show the simultaneously measured time-domain response. 
Remarkably, although the spectral line-width has increased significantly, the ringdown response decays on a timescale comparable to the dataset at a lower power which showed \Qs~$\sim 6250$. Fitting the data to the model, we find \Qr $\sim 6140$, a value very similar to that from the lower power dataset. Although increasing the drive power significantly increases the spectral line-width, the observed value of \Qr~shows that dissipation is unchanged. The observation of \Qr $\gg$ \Qs~demonstrates the importance of dephasing and decoherence in the dynamics of the carbon nanotube motion. We also observe this difference in spectral \Qs~and ringdown \Qr~at different gate voltages (see Supplementary Fig. 11)

\vspace{10pt}

{\bf \Large Discussion}
\\

In the observations presented here, dephasing appears with increased amplitude of the mechanical motion, indicating a non-linear dephasing mechanism. We note that a similar observation of an increasing spectral line-width with increased driving power was recently reported as an indication of nonlinear dissipation in graphene resonators at low temperature\cite{eichler2011nonlinear}. Our results here show that such effects in the spectral quality factor \Qs~can also arise from a power dependent dephasing mechanism, and that fast time-domain measurements such as those presented here are able to distinguish between the two.

Having established the presence of non-linear dephasing in our device, it is interesting to consider what mechanisms could lead to such an effect. 
One possible source is an excess voltage noise on our gate: 
since the mechanical frequency is gate voltage dependent, noise on the gate would give rise to random fluctuations in mechanical frequency, which would increase \Qs~through spectral (inhomogeneous) broadening. 
From the dispersion of the mechanical frequency with gate voltage, we estimate that a gate voltage noise of 45~mV would be needed to produce the observed spectral broadening. This corresponds to a gate voltage noise larger than the Coulomb peak spacing (see Supplementary Fig. 12) . This is significantly higher than the noise level in our setup, suggesting we can rule out gate voltage noise as the origin of the observed effects. 
Another possibility is effects from the relatively weak Coulomb blockade in our device, in which a fluctuating force from the tunnelling of single electrons could dephase the mechanical motion\cite{meerwaldt2012probing}.
Although we did not observe any strong dependence of \Qs~as the gate voltage was swept across the weak Coulomb blockade features, future experiments at lower temperature where Coulomb blockade effects become more significant could shed light on this mechanism.
A third possible source is dephasing from coupling to the stochastic motion of other mechanical modes\cite{barnard2012fluctuation,venstra2012strongly,PhysRevB.80.174103}.
One way in which mode coupling could explain the amplitude dependent spectral broadening is through an increased heating of the other modes from the larger driving.
These effects could be explored in future experiments through detailed temperature dependence studies.
In very recent work\cite{zhang2014interplay}, changes in the thermomechanical noise
spectrum of a carbon nanotube resonator were seen in response to
non-linear driving forces. Such increases in the thermomechanical noise (heating) from large driving amplitudes could also play a role. It is an interesting question if the effects observed in that work could play a role in the spectral- and
time-domain mechanical response studied here

In summary, we have used a high-speed readout technique to measure the mechanical ringdown of a carbon nanotube. 
Using this technique, we demonstrate decoherence of the mechanical motion from an amplitude-dependent dephasing process.
Future studies of motion in the time domain could potentially identify the origin of this dephasing and explore dissipation and decoherence in carbon nanotube motion.\\
\clearpage

\noindent
\textbf{Acknowledgments}\\
We would like to thank H.S.J. van der Zant for useful discussions.
The work was supported by FOM/NWO/OCW and the European Union's Seventh Framework Programme (FP7) under grant agreement no. 318287,  project LANDAUER.\\

\noindent
\textbf{Author contributions}\\
B.H.S, performed the experiments; B.H.S. and H.B.M. fabricated the samples; B.H.S., V.S., W.J.V. and G.A.S. wrote the manuscript; all authors discussed the results and contributed to the manuscript.\\


\end{document}





%
%
%
\newcommand{\Tr}{$T_\text{1}$}
\newcommand{\Ts}{$T_\text{2}$}
\newcommand{\Qr}{$Q_\text{R}$}
\newcommand{\Qru}{$Q_\text{Ru}$}
\newcommand{\Qs}{$Q_\text{S}$}
\newcommand{\Qd}{$Q_\text{D}$}
\newcommand{\Vm}{$V_\text{M}$}
\newcommand{\Vb}{$V_\text{B}$}
\newcommand{\Vg}{$V_\text{G}$}
\newcommand{\Vgm}{$V_\text{G}^\text{M}$}
\newcommand{\Ic}{$I_\text{C}$}
\newcommand{\Ib}{$I_\text{B}$}
\newcommand{\fd}{$f_\text{d}$}
\newcommand{\fp}{$f_\text{p}$}
\newcommand{\degree}{\ensuremath{^\circ}}

%
%

\title{Supplementary Information}
\maketitle

\begin{center}
\begin{figure}[ht]
\centerline{\includegraphics[width=0.8\linewidth]{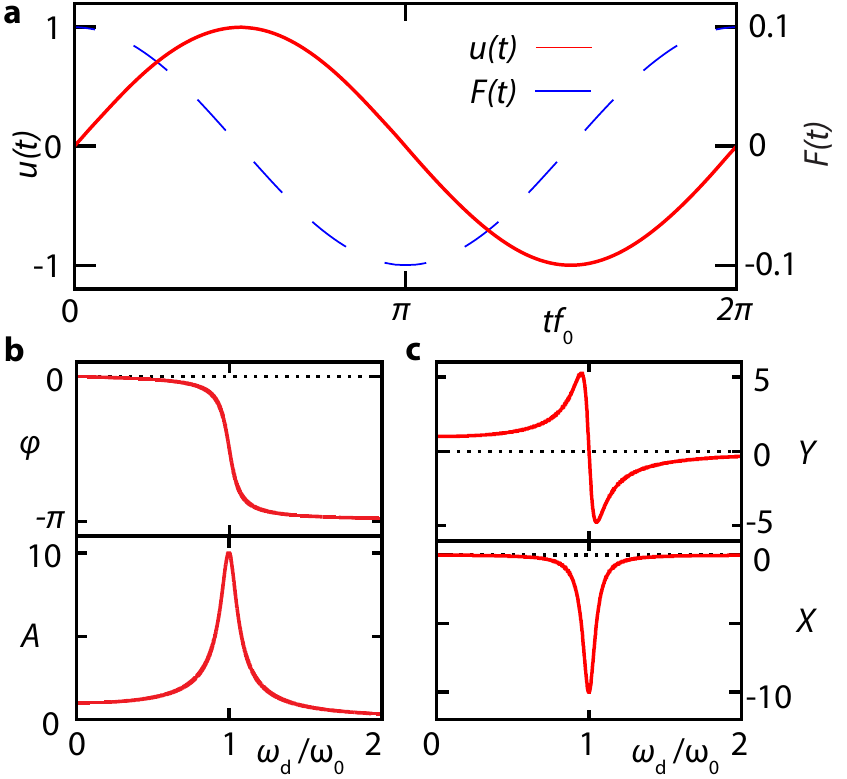}}
\caption{\label{rd:eqhh0}
\textbf{Calculated time and spectral response of a linear resonator with $Q_\text{R} = 10$.}
\textbf{a} The solid red line shows the displacement of the resonator $u(t)$ (Eq. \ref{rd:eqSS}). 
The blue dashed line shows the continuous-driving force $F(t) = \cos( \omega_d t)$, when driven on resonance.
\textbf{b} Phase ($\phi$, top panel), and magnitude ($A$, bottom panel of the response, as a function of the drive frequency.)
\textbf{c} real ($Y$, top panel), and imaginary ($X$, bottom panel) part of $H_{HO}(\omega_d)$.}
\end{figure}
\clearpage

\begin{figure}[ht]
\centerline{\includegraphics[width=0.8\linewidth]{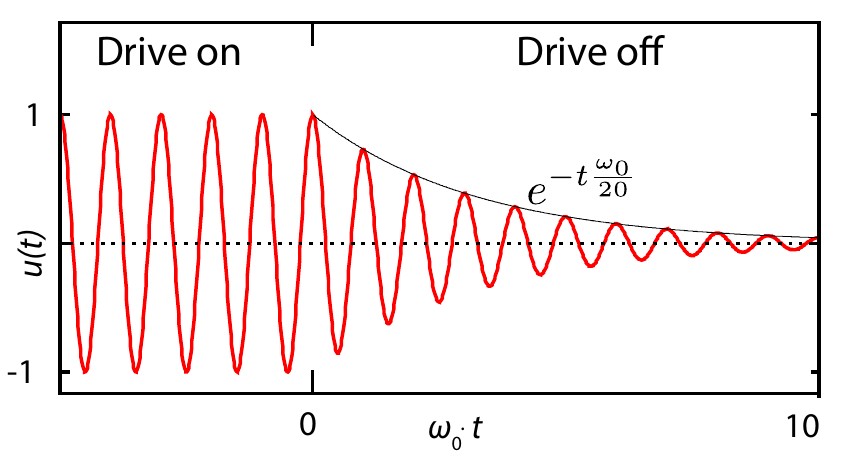}}
\caption{\label{rd:switch} \textbf{Ringdown of the displacement of a linear resonator.} The resonator is continuously driven at resonance until $t = 0$ at which the driving is switched off.
From that time onwards the amplitude of the resonator decays while it is still resonating at its resonance or eigen frequency $\omega_0$.}
\end{figure}
\clearpage

\begin{figure}[ht]
\centerline{\includegraphics[width=0.8\linewidth]{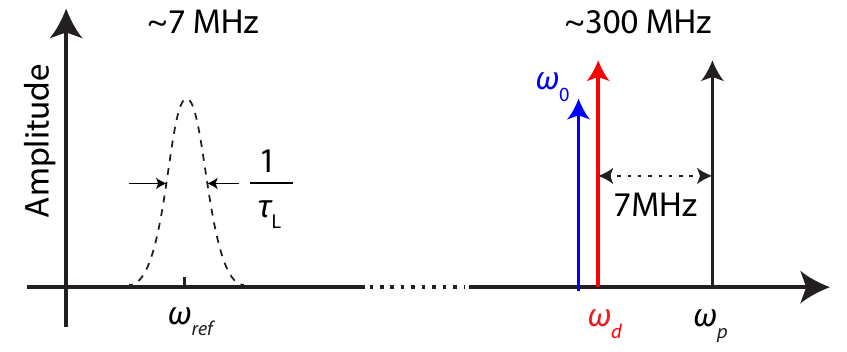}}
\caption{\label{rd:freqan} \textbf{Frequency analysis of the lock-in measurement scheme.}
The lock-in detects frequencies near $\omega_{ref}$ which is chosen to be at 7~MHz.
The probe frequency $\omega_{\text{p}}$ is detuned by 7~MHz from the drive frequency $\omega_{\text{d}}$.
The resonance frequency of the resonator ($\omega_0$, blue arrow) is typically 300~MHz.}
\end{figure}
\clearpage

\begin{figure*}[ht]
\centerline{\includegraphics[width=0.8\linewidth]{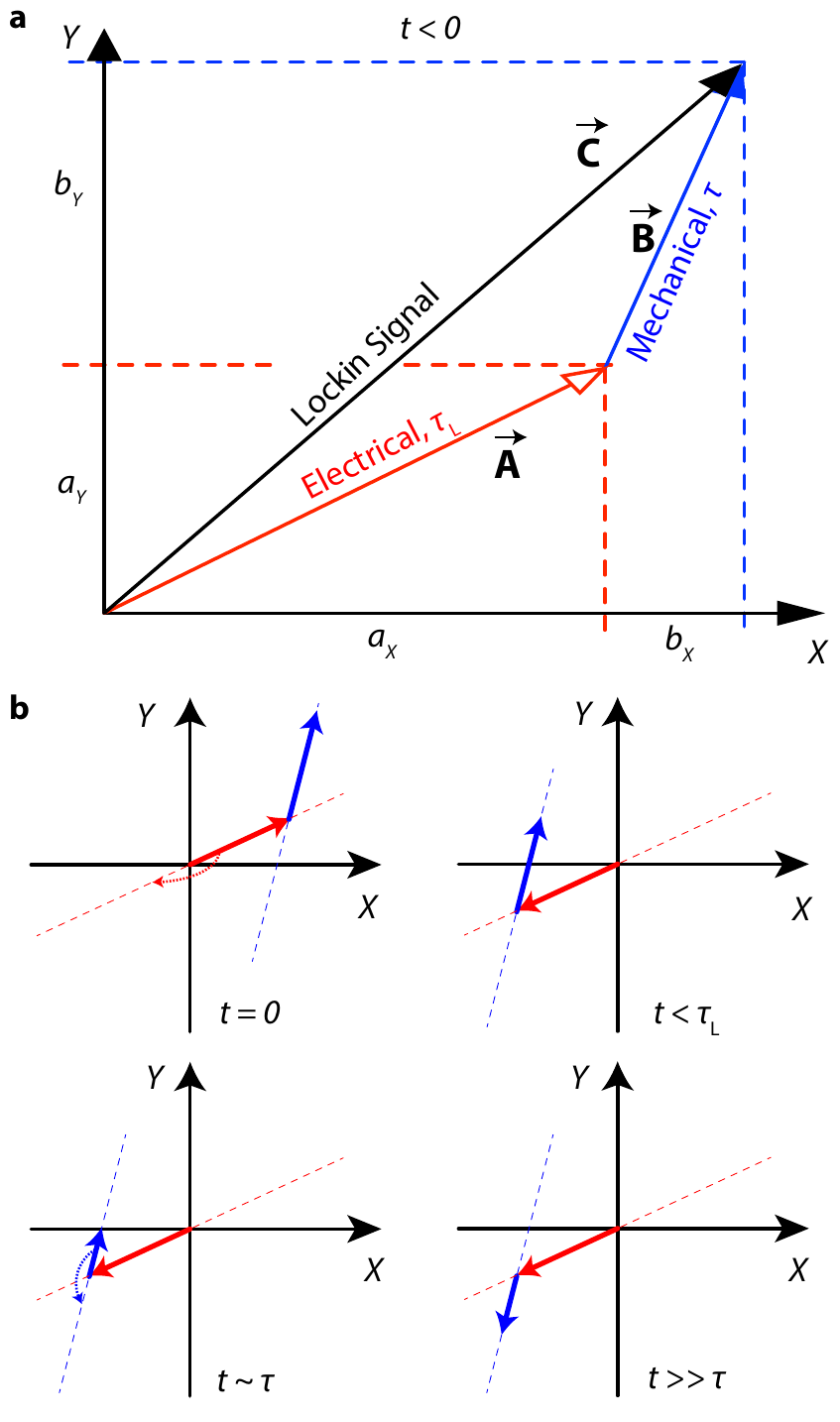}}
\caption{\label{rd:vec} 
\textbf{Vector diagram of the mechanical and electrical signal composition}, 
\textbf{a,b} The detected signal at the lock-in amplifier is represented as a vector $\vec{\textbf{C}}$.
$\vec{\textbf{A}}$ represents the electrical signal and decays with the lock-in time constant $\tau_L$.
$\vec{\textbf{B}}$ represents the mechanical signal which decays with the mechanical ringdown time constant $\tau$.}
\end{figure*}
\clearpage

\begin{figure}[ht]
\centerline{\includegraphics[width = 0.8\linewidth]{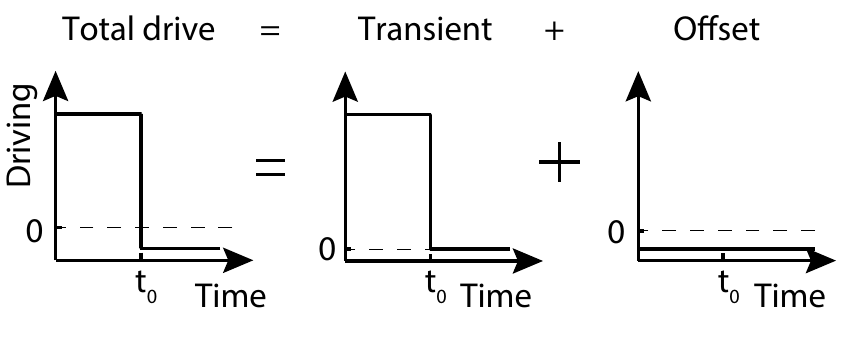}}
\caption{\label{rd:noise}
\textbf{Schematics showing the decomposition of the driving amplitude in the presence of a residual driving signal.}
From left to right: 
In the presence of a residual driving signal with a $180^{o}$ phase shift, the signal drops to a negative offset, after turing the switch off.
This can be decomposed into two components:
One is the transient regime, when the driving is switched off, it drops to zero.
The other component is a constant negative offset.
For a linear resonator the displacement $u(t)$ is given by the transient response of an ideal step function summed with the time-independent effective offset.}
\end{figure}
\clearpage

\begin{figure}[ht]
\centerline{\includegraphics[width = 0.8\linewidth]{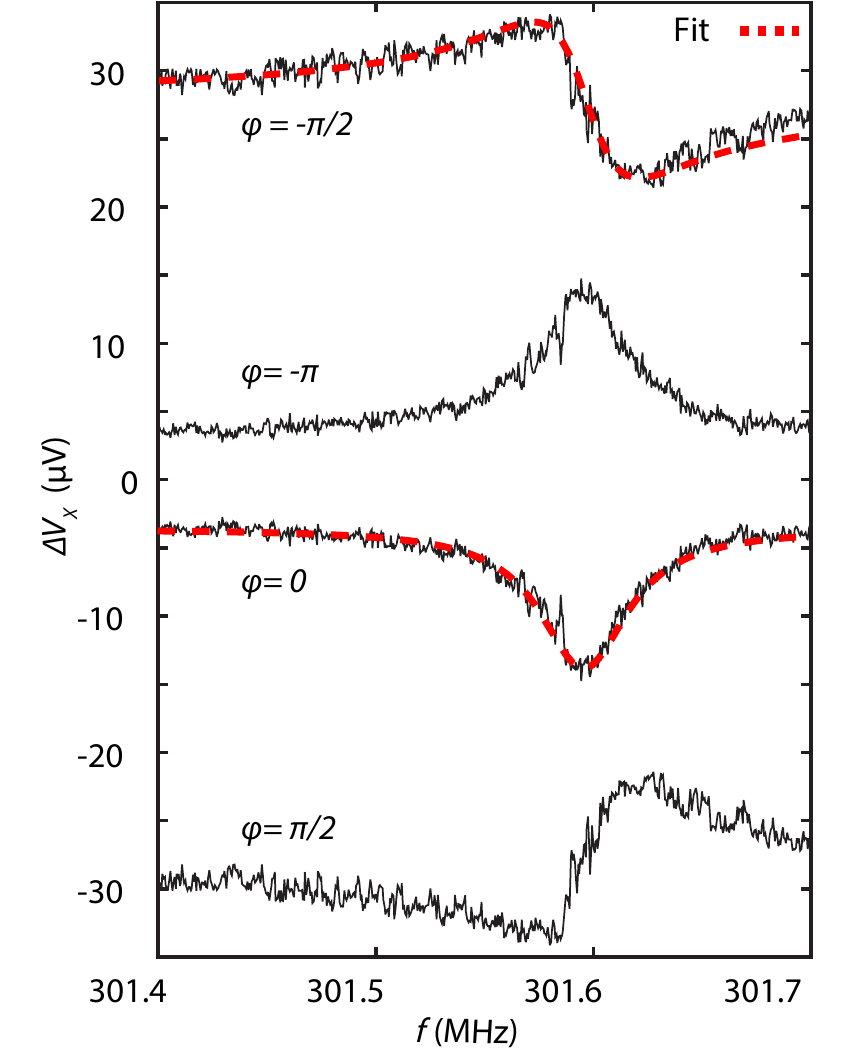}}
\caption{\label{rd:phase} \textbf{Signal line shape as a function of phase rotation}. 
The signal was recorded at a fixed gate voltage \Vg = 0.4~V and a fixed bias voltage \Vb~= -5~mV.
The dashed red curve is a fit to the data yielding \Qs~= 6192.
(The rotation was done using Eq. \ref{rd:eq_rot}).
}
\end{figure}
\clearpage

\begin{figure}[ht]
\centerline{\includegraphics[width = 0.8\linewidth]{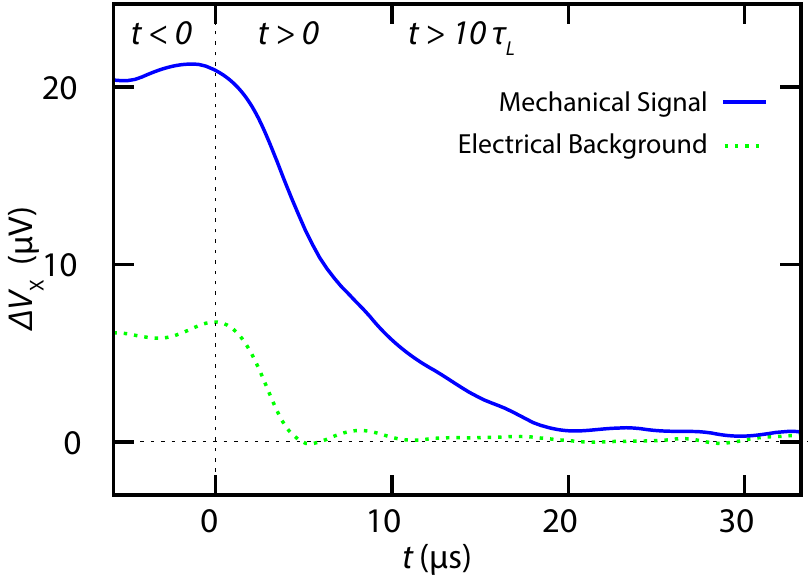}}
\caption{\label{rd:meas1}
\textbf{Mechanical ringdown response of the carbon nanotube resonator.} 
Measured response of the rotated quadrature versus time.
The gate voltage is fixed at \Vg~ = 0.4~V, 
and the phase has been rotated to ($\phi = 0$) such that $\Delta V_X$ represents the amplitude.
The blue solid curve is obtained when the resonator is driven on-resonance ($\omega_d = \omega_0 = 301.596$~MHz). The  green dotted curve represents the signal obtained when the driving force is far off-resonance ($\omega_d=301.4$~MHz). 
The detuning is large enough such that the green dashed curve consists of only the electrical mixing signal. 
The lock-in time constant is 900~ns.}
\end{figure}
\clearpage

\begin{figure*}[ht]
\centerline{\includegraphics[width = 0.8\linewidth]{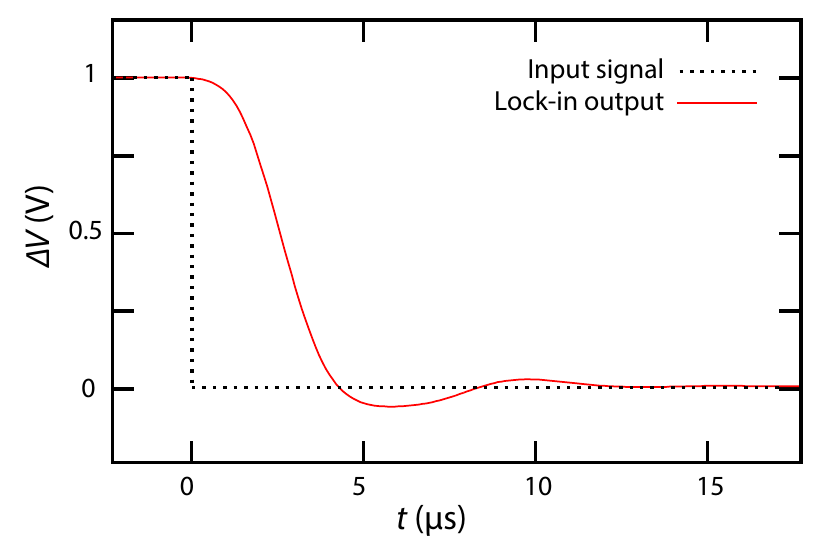}}
\caption{\label{rd:lock-inresponse} \textbf{Measurement of lock-in filter function.} Black dashed line: Input signal into the lock-in (before mixing with lock-in ref. frequency). Red line: output signal from the lock-in.
The lock-in filter function is equal to the derivative of the output signal (see Supplementary Note 3).
For this measurement, $\tau_L = 900$~ns.}
\end{figure*}
\clearpage

\begin{figure}[ht]
\centerline{\includegraphics[width = 0.8\linewidth]{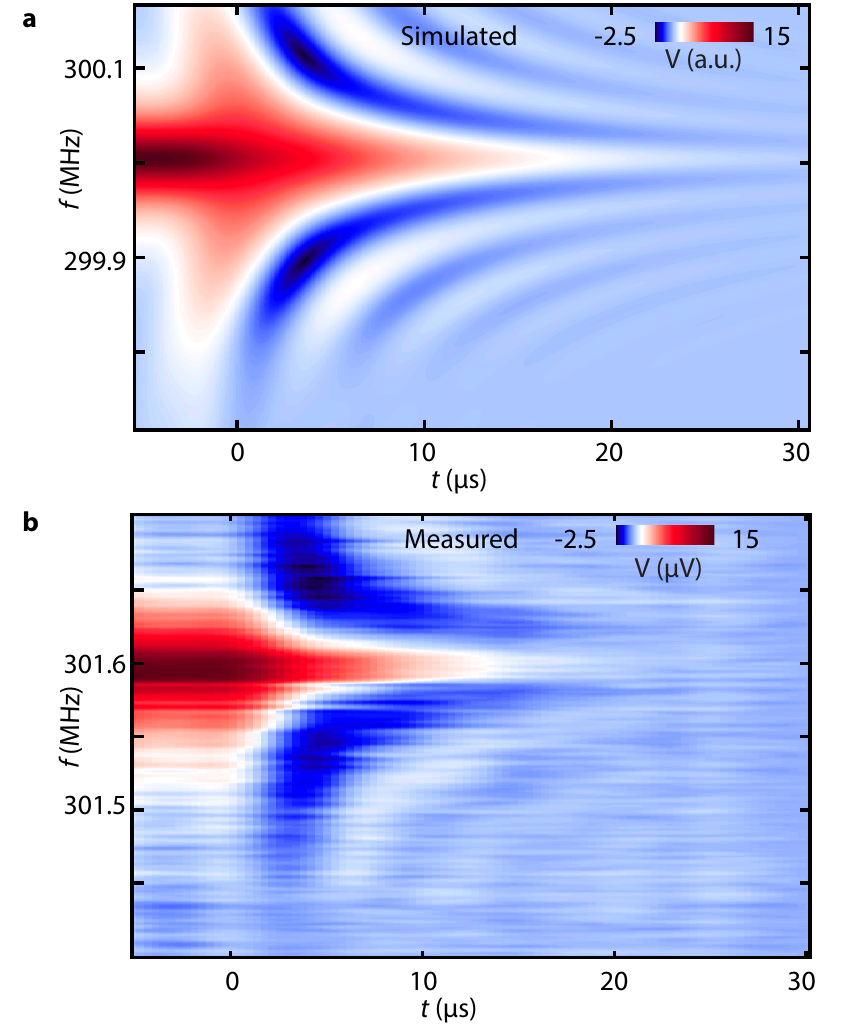}}
\caption{\label{rd:simdataset} \textbf{Simulated and measured voltage from the lock-in.} 
The shown signal corresponds to the response magnitude.
\textbf{a} Simulated results. (using \Qs~= 1410 and \Qr~= 6140)
\textbf{b} Measured data set (\Qs~= 1410, \Qr~= 6140) (from which Fig.3 in the main text is extracted).}
\end{figure}
\clearpage

\begin{figure}[ht]
\centerline{\includegraphics[width = 0.8\linewidth]{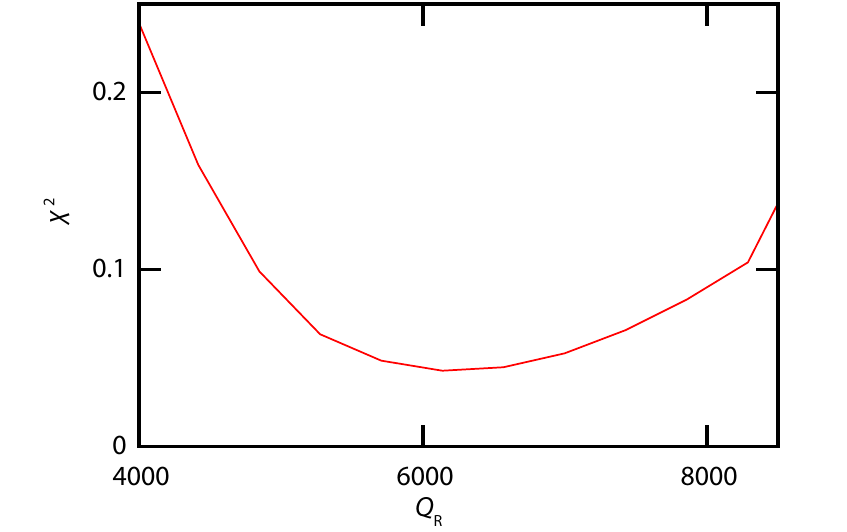}}
\caption{\label{rd:simdataset2}
\textbf{Mean square deviation between $\chi^2$ between the measured and the simulated data set as a function of \Qr}
This is the fitting result for the data set presented Fig. 4b of the main text.
The spectral quality factor is \Qs~= 1407, the fitted ringdown is found to be around \Qr~$\sim 6140$.}
\end{figure}
\clearpage

\begin{figure*}[ht]
\centerline{\includegraphics[width= 0.8\linewidth]{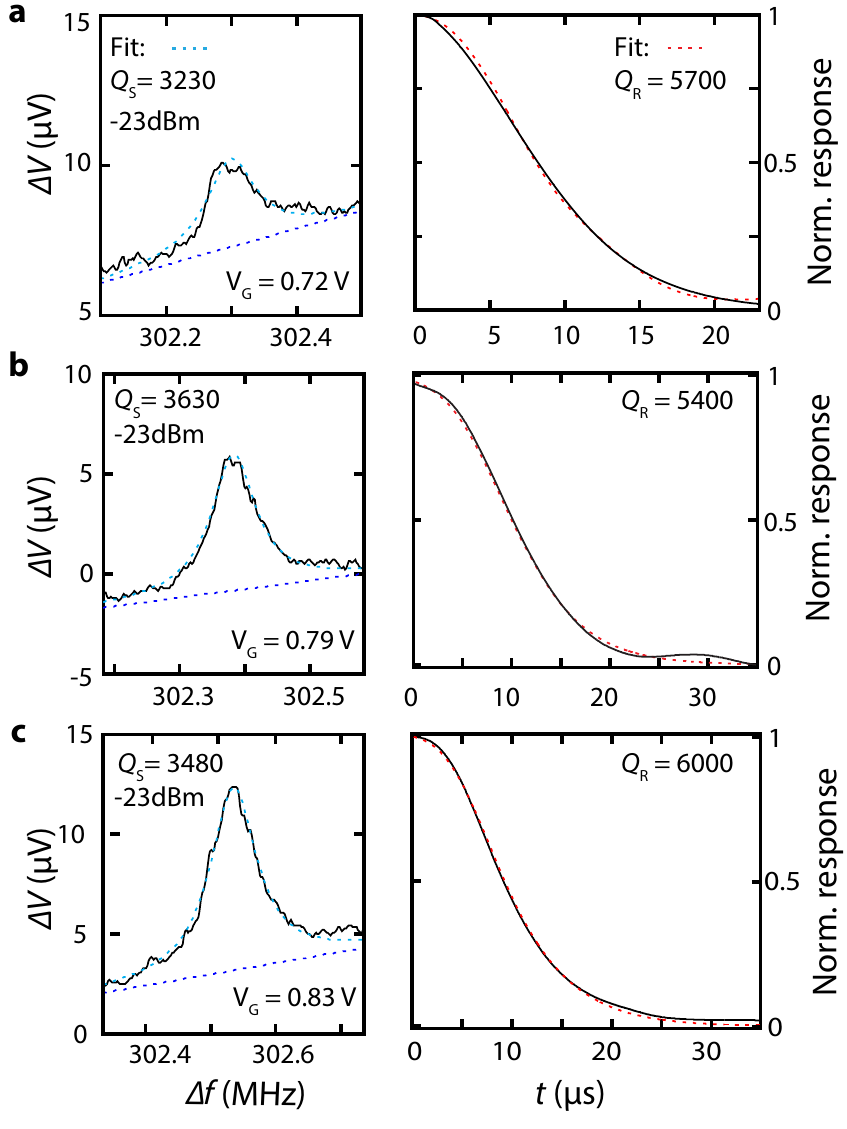}}
\caption{\label{rd:diffvgs}
\textbf{Decoherence due to dephasing at three different gate voltages \Vg, at  $T$ = 3.1~K.} 
Left and right panels show the spectral and normalised time-domain response of the resonator.
The errorbars on \Qs~and \Qr~are $\pm 100$ and $\pm 600$ respectively (with $5\%$ change in $\chi^2$).}
\end{figure*}
\clearpage

\begin{figure*}[ht]
\centerline{\includegraphics[width= 0.8\linewidth]{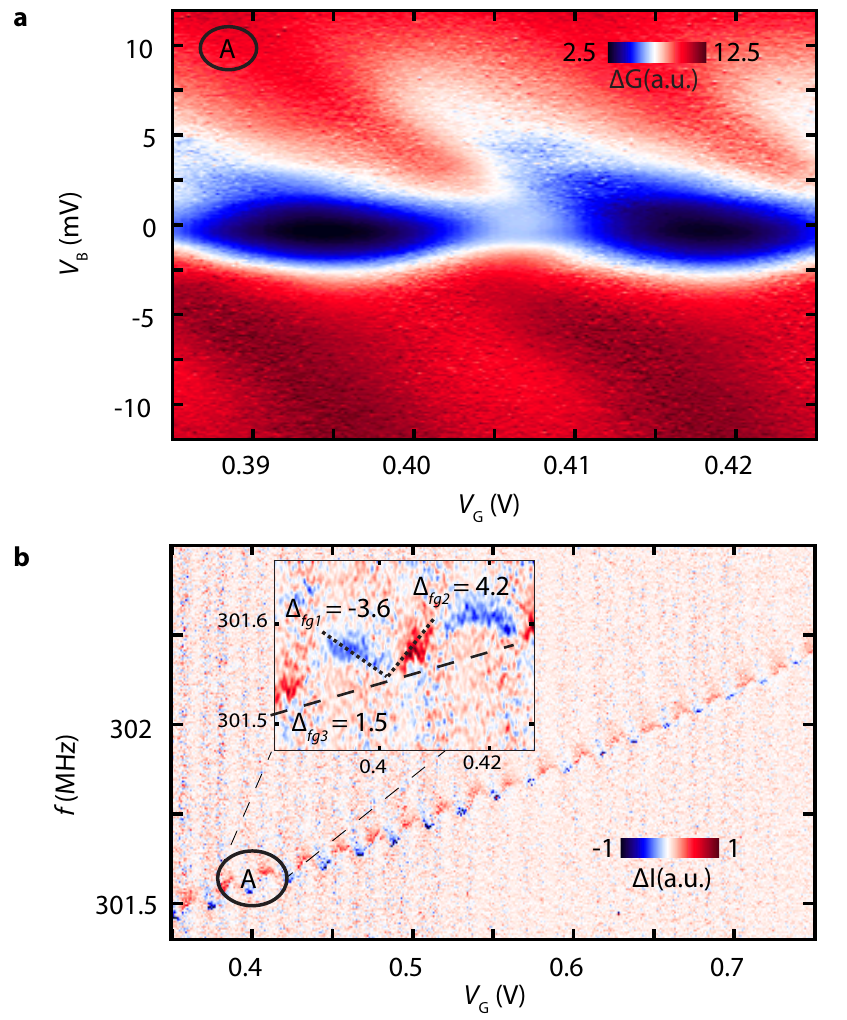}}
\caption{\label{rd:A}
\textbf{Electronic and mechanical response of the CNT at $T$ = 2~K.} 
\textbf{a} Colour-scale-plot of the measured differential conductance as a function of \Vg~and \Vb~measured across two Coulomb peaks.
\textbf{b} Measured change in rectification current (colour) as a  function of gate voltage and drive frequency at \Vb~= -5~mV. 
The inset is a magnification of region A, taken by a second measurement.
Dashed lines and the values $\Delta_{fg1} = -3.6~\text{MHz} ~ \text{V}^{-1}$, $\Delta_{fg2}= 4.2~\text{MHz} ~ \text{V}^{-1}$ and $\Delta_{fg3} = 1.5~\text{MHz} ~ \text{V}^{-1}$ indicate the slopes of the frequency change as a function of \Vg~.}
\end{figure*}
\end{center}

\clearpage

\section*{Supplementary Note 1. Theory on ringdown of a mechanical resonator}

\label{rd:section:theory}
The CNT is modelled as a damped driven harmonic oscillator with an equation of motion\cite{poot2012mechanical}:
\begin{eqnarray}
m \ddot{u} = -k_R u - m \dot{u} ~ \tau^{-1} + F(t) \label{rd:eqm}~,
\end{eqnarray}
where $m$ is the mass of the CNT with a displacement $u$ relative to the equilibrium position, 
a spring constant $k_R = m \omega_0^2$, a decay time $\tau$ and a driving force $F(t)$.
By taking the Fourier transform, $\mathfrak{F}[x(t)] = \int_{-\infty}^{\infty} x(t) \exp(-i\omega_d t) dt $ of the equation of motion (Eq. \ref{rd:eqm}) and taking a harmonic driving force $F(t) =  F_0 \cos(\omega_d t)$, the following transfer function is obtained:
\begin{eqnarray}
H_{HO}(\omega_d) = k_R \frac{u(\omega_d)}{F(\omega_d)} = \frac{\omega_0^2}{\omega_0^2-\omega_d^2 + i \omega_d \omega_0 /Q_R} \label{rd:eqHHO} ,
\end{eqnarray}
where $\omega_d = 2 \pi f_d$ and the quality factor is related to the decay time as $Q_R = \tau \omega_0$ .\\

\textbf{Driven response.}
In continuous driving with a driving force $F(t) = F_0 \cos(\omega_d t)$ 
a steady state solution for the equation of motion is:
\begin{eqnarray}
u(t) = X_d~\cos(\omega_d t) + Y_d~\sin(\omega_d t) \label{rd:eqSS}~,
\end{eqnarray}
where $X_d = A\cdot \cos(\phi)$, $Y_d =-A\cdot \sin(\phi)$, $A = |H_{HO}(\omega_d)|$ and $\phi = \angle H_{HO}(\omega_d)$ is a phase angle of the response function, which gives the relative phase between the nanotube motion and the driving force.
Supplementary Fig. \ref{rd:eqhh0} shows time and spectral responses of a linear resonator.
In Supplementary Fig. \ref{rd:eqhh0}a the resonator is driven at resonance frequency.
The displacement $u(t)$ lags behind the driving force $F(t)$ by a phase difference of $-\pi/2$.
This phase offset between drive and displacement of the resonator depends on the drive frequency $\omega_d$.
In the top panel of Supplementary Fig. \ref{rd:eqhh0}b the phase lag of the displacement with respect to the actuation frequency is plotted; the bottom panel shows the amplitude response of the resonator.
With changing drive frequency the phase $\phi$ changes from 0 to $-\pi$ across its resonance position.
At resonance the amplitude $A$ has its maximum while the phase lags behind the driving force by $-\pi/2$.\\

\textbf{Switching the drive off.}
In this section we will consider a harmonic oscillator that is driven by a constant driving force $F(t) = F_0 \cos(\omega_d t)$ for $t \leq 0$, and undergoes free evolution with $F=0$ for $t>0$. 
For $t\leq0$, $u(t)$ is given by Eq. \ref{rd:eqSS}. 
For $t>0$, using the equation of motion (Eq. \ref{rd:eqm}) with $F=0$ and assuming $Q > 1$, $u(t)$ is given by the following equation;
\begin{eqnarray}
u(t) =\left[ X_0 \cos(\omega_0 t) + Y_0 \sin(\omega_0 t) \right] e^{-\frac{t}{2 \tau}} \label{rd:eq2}.
\end{eqnarray}
The constants $X_0$ and $Y_0$ are determined by matching the position $u$ and velocity $\dot u$ of the two solutions at $t=0$, giving $X_0= X_d$ and $Y_0 = Y_d \frac{\omega_d}{\omega_0}$. 
Supplementary Fig. \ref{rd:switch} shows the displacement evolution $u(t)$ of a linear resonator with time.
The resonator is driven with a resonant driving force in (Supplementary Fig. \ref{rd:switch}a) and then switched off at $t = 0$.
It is interesting to note that for $t>0$ the displacement of the resonator $u(t)$ oscillates at its natural frequency $\omega_0$ independent of the initial driving frequency $\omega_d$.
Supplementary Fig. \ref{rd:switch} shows the ringdown a resonator driven driven at $\omega_d=\omega_0$ for $t\leq0$.
In this example, the decay time $\tau = 10/ \omega_0$ and $Q = \tau \omega_0 = 10$. 
For $t\leq0$, $u(t)$ oscillates with constant amplitude and for $t>0$ the amplitude of the oscillations decays with a time constant $\tau = Q/ \omega_0$.
To summarise, the equations describing the mechanical motion of the resonator at any given drive frequency can be written as:
\begin{eqnarray}
u(t) = 
\begin{cases}
t \leq 0, &  X(t)\cos(\omega_d t) + Y(t)\sin(\omega_d t)\\ 
t > 0, &  X(t) \cos(\omega_0 t) + Y(t) \sin(\omega_0 t),
\end{cases} \label{rd:eq3xxx}
\end{eqnarray}
where $X(t)$ and $Y(t)$ are given by:
\begin{eqnarray*}
X(t) &= 
\begin{cases}
t \leq 0 , 
&X_d, \\
t > 0 , 
&X_d \cdot e^{-\frac{t}{2 \tau}}, 
\end{cases}\\
Y(t) &= 
\begin{cases}
t \leq 0 , 
&Y_d  \\
t > 0 , 
&Y_d \frac{\omega_d}{\omega_0}  \cdot e^{-\frac{t}{2 \tau}}.
\end{cases}
\end{eqnarray*}

\clearpage

\section*{Supplementary Note 2. Electrical transduction of the mechanical response from the carbon nanotube resonator.}
To measure the mechanical response of the carbon nanotube, we use a variant of a two source mixing technique. 
We apply two RF signals with frequencies $\omega_d$ and $\omega_p$ to the gate electrode.
The signal at $\omega_d$ is used to drive the mechanical resonator and the signal at $\omega_p$ is used to probe the mechanical response.
The signal $\omega_p$ is detuned from $\omega_d$ by an amount $\omega_{ref} = \omega_p - \omega_d$. We have chosen $\omega_{ref} \sim 2 \pi \times 7~$MHz, which is much larger than the mechanical line width. Therefore the CNT will not be driven by the probe signal at $\omega_p$.
We will now consider the transduction of the mechanical motion, given by eq. \ref{rd:eq3xxx}, 
into a signal that is detected in the experiment.
The total electrostatic voltage $\delta V_\text{G}^{e}$ is given by:
\begin{eqnarray}
\delta V_\text{G}^{e} &=
\begin{cases}
t \leq 0, &
V_\text{G}^{ac,d} \cos(\omega_d t) 
+ V_\text{G}^{ac,p} \cos(\omega_p t) \\
t > 0, & 
V_\text{G}^{ac,p} \cos(\omega_p t)
\end{cases}
\end{eqnarray}
where $V_\text{G}^{ac,d}$ and $V_\text{G}^{ac,p}$ are the amplitude of the drive and probe signals, respectively.
The effect of the mechanical motion can be captured by considering an effective oscillating gate voltage given by:
\begin{eqnarray}
\delta V_\text{G}^{m}& = & 
\frac{V_\text{G} }{C_\text{G}} \frac{dC_\text{G}}{du} 
\cdot u(t)
\label{jem:eq:ac32}, 
\end{eqnarray}
where $C_\text{G}$ is the capacitance between CNT and the local gate.
It is important to note that $\delta V_\text{G}^{m}$ has a frequency of 
$\omega_d$ for $t \leq 0$,
and  $\omega_0$ for $t > 0$.
To analyse the electrical signals generated by the carbon nanotube we can treat it as an electrical mixing experiment which now includes an oscillating gate voltage:
\begin{eqnarray}
\delta V_\text{G} =  \delta V_\text{G}^{e} + \delta V_\text{G}^{m}. \label{rd:dvg}
\end{eqnarray}
The displacement of the CNT is represented by $u(t)$.
When we assume that the CNT is a linear resonator, the displacement $u(t)$ is  given by Eq. \ref{rd:eq3xxx}.

Now, we will consider the generic case when the current through the nanotube is an arbitrary function of the gate voltage $I=I(V_\text{G})$.
By doing a Taylor expansion of the current through the CNT with gate voltage up to the second order, we get:
\begin{eqnarray}
I(\delta V_G) &=& I_{0} + \frac{dI}{d V_G}\delta  V_G  + \frac{1}{2}\frac{d^2I}{d V_G^2}\left(\delta  V_G\right)^2,
\label{rd:eq:current}
\end{eqnarray}
where $I_{0}$ is the dc current.
We now need to consider what will be measured by the lock-in amplifier.
The lock-in measures only components of the signal 
that are within the lock-in measurement bandwidth ($\frac{1}{\tau_{L}}$) 
around the reference frequency ($\omega_{ref}$), as illustrated in Supplementary Fig. \ref{rd:freqan}.
The lock-in is not sensitive to DC currents, therefore we can neglect the first term.
Also, since $\omega_0$, $\omega_d$, $\omega_p >> \omega_{ref} =\omega_p - \omega_d $, the second term in equation \ref{rd:eq:current} will not give any signal, that will be detected in the experiment.
The third term includes products of cosines and sines and therefore can introduce mixed down signals detected by the lock-in.
By substituting Eq. \ref{rd:dvg} into the third term, we get:
\begin{eqnarray}
\Delta I &=& \frac{1}{2}\frac{d^2I}{d V_G^2} 
\left[
(\Delta V_\text{G}^{e})^2 + 2 \delta V_\text{G}^{m} \delta V_\text{G}^{e} + (\delta V_\text{G}^{m})^2
\right] 
\label{rd:eq:10}
\end{eqnarray}
Two contributions give a signal near $\omega_{ref}$: the first is from the term $(\delta V_\text{G}^{e})^2$, which will give an electrical background from the electrical mixing of the probe and drive voltages.
The second term $2 \delta V_\text{G}^{m} \delta V_\text{G}^{e}$  gives us an AC current near $\omega_{ref}$.
The third term $(\delta V_\text{G}^{m})^2$ will not contribute to the lock-in signal as it only contains DC and $\sim 2\omega_0$ frequencies. (This is the term is used in previous DC rectification experiments.\cite{huettel2009carbon})
Thus, the total current that contributes to the signal detected by the lock-in is given by:
\begin{eqnarray}
\Delta I_L &= \frac{1}{2}\frac{d^2I}{d V_G^2} 
\begin{cases}
t \leq 0, &
(\Delta V_\text{G}^{e})^2 + 2 \delta V_\text{G}^{m} \delta V_\text{G}^{e} \\
t > 0, & 
2 \delta V_\text{G}^{m} \delta V_\text{G}^{e}.
\end{cases}
\end{eqnarray}\\

\textbf{Quadrature signal during ringdown.}
We start by considering a signal without any electrical background $(\Delta V_\text{G}^{e})^2$, which is not present when the resonator is ringing down.
For $t > 0$, $\Delta I_L = \Delta I_{L}^{mech}$  which is the signal measured by the lock-in, generated by the mechanical motion of the carbon nanotube:
\begin{eqnarray}
\Delta I_{L}^{mech} &=& \frac{d^2I}{d V_G^2} \delta V_\text{G}^{m} \delta V_\text{G}^{e} \\
\delta V_\text{G}^{e} &=& V_\text{G}^{ac,p} \cos(\omega_p t)\\
\delta V_\text{G}^{m} &=&
\alpha \cdot u(t) \\
\alpha &=& \frac{V_\text{G} }{C_\text{G}} \frac{dC_\text{G}}{du},
\end{eqnarray}
where $\alpha$ is a factor which translates the displacement of the carbon nanotube into a voltage.
The input signal at the lock-in becomes:
\begin{eqnarray}
\Delta I_{L}^{mech} &=& \frac{d^2I}{d V_G^2} 
\alpha \left[
X(t) \cos(\omega_0 t) + Y(t)\sin(\omega_0 t)
\right] 
\cdot
V_\text{G}^{ac,p} \cos(\omega_p t).
\label{rd:eq:12}
\end{eqnarray}

This signal is then filtered by the low pass filter before it enters the lock-in.
Retaining only the low frequency components ($\Delta \omega = \omega_p - \omega_0$), Eq. \ref{rd:eq:12} becomes:
\begin{eqnarray}
&\Delta I_{L}^{mech} = 
&\frac{1}{2}
\frac{d^2I}{d V_G^2} V_\text{G}^{ac,p}
\alpha \left[
X(t) \cos(\Delta \omega t) + Y(t)\sin(\Delta \omega t)
\right] 
\label{rd:eq:14}
\end{eqnarray}

This signal is at frequency $\Delta \omega$ and is read out at the lock-in with respect to the reference signal $\omega_{ref}=\omega_p-\omega_d$.\\

The two outputs of the lock-in quadratures for a time $t > 0$ with $\omega_{ref} - \Delta \omega = \omega_0-\omega_d = \omega_{beat}$ are the following:
\begin{eqnarray}
X_{L}^{mech} &=& \frac{1}{2} \frac{d^2I}{d V_G^2} V_\text{G}^{ac,p} \alpha 
\left[
X(t) \cos\left(\omega_{beat} t \right) +
Y(t) \sin\left(\omega_{beat} t \right)
\right]
\label{rd:eq:xmech}\\
Y_{L}^{mech} &=& \frac{1}{2} \frac{d^2I}{d V_G^2} V_\text{G}^{ac,p} \alpha 
\left[
X(t) \sin\left(\omega_{beat} t \right) +
Y(t) \cos\left(\omega_{beat} t \right)
\right].
\label{rd:eq:ymech}
\end{eqnarray}
When driven on resonance $(\omega_d = \omega_0,~ \omega_{beat} = 0)$, $\cos( \omega_{beat} t) = 1$ and $\sin(\omega_{beat}) = 0$. The quadrature outputs of the lock-in represent the time-dependent amplitudes $X(t)$ and $Y(t)$.\\

For $t > 0$, the driving signal is switched off.
Now $\omega_d$ is not necessarily equal to $\omega_0$.
In this case the quadrature outputs from the lock-in are:
\begin{eqnarray}
X_{Lr}^{mech} &=& \frac{1}{2} \frac{d^2I}{d V_G^2} V_\text{G}^{ac,p} \alpha 
\left[
X_d \cos\left(\omega_{beat} t \right) +
Y_d \frac{\omega_d}{\omega_0} \sin\left(\omega_{beat} t \right)
\right]
e^{-\frac{t}{2 \tau}} \label{rd:eq:ymech244444} \\
Y_{Lr}^{mech} &=& \frac{1}{2} \frac{d^2I}{d V_G^2} V_\text{G}^{ac,p} \alpha 
\left[
X_d \sin\left(\omega_{beat} t \right) +
Y_d \frac{\omega_d}{\omega_0}  \cos\left(\omega_{beat} t \right)
\right]
e^{-\frac{t}{2 \tau}}.
\end{eqnarray}\\

\textbf{Quadrature signals when drive is on.}
For $t \leq 0$, the resonator is driven. 
The frequency of the motion is the same as the drive frequency $\omega_0 = \omega_d$.
The lock-in quadratures (without any electrical mixing $(\Delta V_\text{G}^{e})^2$), then simplify to: 
\begin{eqnarray}
X_{Ld}^{mech} &=& \frac{1}{2}\frac{d^2I}{d V_G^2} V_\text{G}^{ac,p} \alpha 
X_d
\label{rd:eq:xmech2}\\
Y_{Ld}^{mech} &=& \frac{1}{2}\frac{d^2I}{d V_G^2} V_\text{G}^{ac,p} \alpha 
Y_d
\label{rd:eq:ymech2}.
\end{eqnarray}

In addition to the mechanical mixing signals, there is also an electrical mixing signal present when the resonator is driven.
This gives rise to a constant voltage offset, $X_{L}^{elec}$ and $Y_{L}^{elec}$, which are independent of time and 
drive frequency (aside from small overall slopes from weak frequency dependent transmission of the RF cables in the setup):
\begin{eqnarray}
\Delta I_{L}^{elec} &=& \frac{1}{2}\frac{d^2I}{d V_G^2}  V_\text{G}^{ac,p} V_\text{G}^{ac,d}  \cos(\omega_{ref} t)\\
X_{L}^{elec} &=&\frac{1}{2}  \frac{d^2I}{d V_G^2} V_\text{G}^{ac,p} V_\text{G}^{ac,d} \label{rd:eq:xelec} \\
Y_{L}^{elec} &=& \frac{1}{2} \frac{d^2I}{d V_G^2} V_\text{G}^{ac,p} V_\text{G}^{ac,d} \label{rd:eq:yelec}
\label{rd:eq:16}
\end{eqnarray}

Combining Eqs. \ref{rd:eq:xmech2} - \ref{rd:eq:16} we obtain:
\begin{eqnarray}
X_L &=
\begin{cases}
t \leq 0, &
X_{Ld}^{mech} + X_{L}^{elec}\\
t > 0, & 
X_{Lr}^{mech} 
\end{cases}\\
Y_L &=
\begin{cases}
t \leq 0, &
Y_{Ld}^{mech} + Y_{L}^{elec}\\
t > 0, & 
Y_{Lr}^{mech} 
\end{cases}.
\end{eqnarray}\

The output of the lock-in amplifier is thus:
\begin{eqnarray}
X_L &=&
\begin{cases}
t \leq 0, &
\beta \cdot
X_d + X_{L}^{elec}\\
t > 0, & 
\beta \cdot
\left[
X_d \cos\left(\omega_{beat} t \right) +
Y_d \frac{\omega_d}{\omega_0} \sin\left(\omega_{beat} t \right)
\right]
e^{-\frac{t}{2 \tau}}
\end{cases}
\label{rd:eq:xmech3}\\
Y_L &=&
\begin{cases}
t \leq 0, &
\beta \cdot
Y_d + Y_{L}^{elec}\\
t > 0, & 
\beta \cdot
\left[
X_d \sin\left(\omega_{beat} t \right) +
Y_d \frac{\omega_d}{\omega_0}  \cos\left(\omega_{beat} t \right)
\right]
e^{-\frac{t}{2 \tau}}
\end{cases}
\label{rd:eq:ymech3}\\
\beta &=& \frac{1}{2} \frac{d^2I}{d V_G^2} V_\text{G}^{ac,p} \alpha
\end{eqnarray}


Supplementary Fig. \ref{rd:vec}a illustrates the two recorded quadratures of the lock-in amplifier while driven on resonance. 
$X_{L}^{elec}$ and $Y_{L}^{elec}$ describe the electrical mixing of the probe and drive signal at the CNT.
This is an electrical background signal which is independent of the motion of the CNT,  indicated by the red arrow in Supplementary Fig. \ref{rd:vec}a. 
$X_{L}^{mech}$ and $Y_{L}^{mech}$ describe the mixing of the mechanical motion with the probe signal.
The amplitude of the signal is proportional to ($|H_{HO}(\omega)|$), which is frequency-dependent and has a maximum on resonance ($\omega = \omega_0$). 
As soon as the drive signal is switched off, $X_{L}^{elec}$ and $Y_{L}^{elec}$ decay over time with the lock-in time constant $\tau_L$.
If the lock-in time constant is smaller than the mechanical time constant $\tau_L <  \tau$ the mechanical signal remains such that $X_{L}^{mech}$ and $Y_{L}^{mech}$ is detected.
From Eq. \ref{rd:eq2}, we see that the amplitude of this signal decays with $e^{-\frac{t}{2 \tau}}$ in time.

Supplementary Fig. \ref{rd:vec}b is a smaller version of (Supplementary Fig. \ref{rd:vec}a) for ($t \leq 0$), shows four different stages in time.
At a time $t = 0$ the drive is switched off; both the electrical (red arrow) and the mechanical signal (blue arrow) are present. The black dotted lines indicate the angle of the vectors.
As time progresses $0 < t < \tau_{\text{L}} $, the electrical signal (red) decays at a much faster rate than the mechanical signal (blue).
From now on the mechanical ringdown of the CNT is recorded by the quadratures of the lock-in. At a time $t >> 10 \tau$, the mechanical signal settles.
Finally, both arrows, for the mechanical and the electrical components are inverted.
This inversion is due to residual driving after the driving is switched off.\\

\textbf{Residual driving when the switch is off.}
\label{rd:cmn}
When switching the drive signal off, we note that there is a small residual AC voltage on the gate and it appears to be $180^{o}$ phase shifted.
We attribute this to insufficient AC grounding in the setup, resulting in a common-mode signal.
As illustrated in Supplementary Fig. \ref{rd:noise}, this results in a residual driving, giving rise to a constant offset in time.
This artefact does not affect the transient response from which we extract the ringdown quality factors,
since the displacement $u(t)$ is a linear combination of both effects.
The motion of the resonator for $t > 0$ becomes:
\begin{eqnarray}
u(t) = (X_0 \cos(\omega_0 t) + Y_0 \sin(\omega_0 t) )e^{-\frac{t}{2 \tau}}
- C( X_d \cos(\omega t) +Y_d \sin(\omega t) ),
\end{eqnarray}\
where $C$ is a constant.\\

\textbf{Spectral measurements}
\label{rd_a:phase_rot}
We first turn to the driven motion in the frequency domain. In Supplementary Fig. \ref{rd:phase} we plot the $X$ and $Y$ quadratures detected by the lock-in as a function of drive frequency $f$. 
The gate voltage is fixed at 0.4~V. 
The plot is obtained by slowly stepping the drive (and probe) signal across the mechanical resonance of the CNT,
while recording the $X$ and $Y$ quadratures of the lock-in output signal.
The data can be fitted to the following equation:
\begin{eqnarray}
f_{sp}(\omega) = 
a |H_{HO}(\omega)|e^{i\phi_m} +(b+c \omega )e^{i\theta} \label{rd:eqHHOabs}~,
\end{eqnarray}
where $\phi_m$ and $\theta$ are the phase from the mechanical response and the electrical background signals respectively.
The real and imaginary parts of $f_{sp}(\omega)$ are fitted to the $X$ and $Y$ quadratures respectively, where a,b and c are free parameters.
The spectral quality factor, resonance position and phase are obtained from this fit.\\

\textbf{Phase rotation of the data.}
In the analysis of the data, it is important that the phase offset between the reference and the input signal
is taken into account. This can be done by introducing a rotation matrix:
\begin{eqnarray}
X1 = X\cos(\phi) - Y\sin(\phi)~, \nonumber \\
Y1 = X\sin(\phi) + Y\cos(\phi)~, \label{rd:eq_rot}
\end{eqnarray}
where X1 and Y1 are the rotated quadratures and $\phi$ is the phase by which the two recorded $X$ and $Y$ quadratures are rotated.
In Supplementary Fig. \ref{rd:phase} the rotated quadratures are shown for four different phase offsets.
The phase is rotated from $\phi = -\pi$ to $\phi = \pi /2$.
When $\phi = 0$ the response is similar to the one of Supplementary Fig. \ref{rd:eqhh0}c ($X$), 
which represents the linear response of a mechanical resonator. 
The corresponding $Y$ quadrature $\phi = -\pi /2$, looks similar to ($Y$) in Supplementary Fig. \ref{rd:eqhh0}c.
This quadrature represents the amplitude of the carbon nanotube motion.
For all measurements we have performed this rotation to correct the phase offset between the reference and the input signals.
After rotation of the $X$ and $Y$ quadratures, the electrical background offset signal can be subtracted.
The convenience of working with quadratures is that one can rotate the phase so that one of the quadratures represents the amplitude of the mechanical resonance, and can be fitted to extract the quality factor (as long as the mechanical resonator is driven in the linear-response regimes, before the onset of Duffing non-linearities).

From the two curves which are rotated to $\phi = 0$ and $\phi = -\pi /2 $ (Supplementary Fig. \ref{rd:phase}) we can determine the resonance frequency and quality factor by fitting the response to $f_{sp}(\omega)$. 
In Supplementary Fig. \ref{rd:phase}, the fit is shown as a red dashed line on top of the data.
We find a resonance frequency $f_{res} = 301.596$~MHz and a spectral $Q$-factor of the resonator \Qs~= 6192.\\ 

\textbf{Estimation of the motional amplitude for the measurement shown in Fig 4 in the main text:}
The amplitude of the mechanical resonator can be estimated from the ratio of the peak voltage at resonance with the background electrical mixing voltage (eq. 13, 14), given by:
\begin{eqnarray}
u =  \frac{V^{m}}{V^{bg}} \frac{V_G^{ac} }{V_{\text{G}}} \cdot h_0 \cdot ln(2 h_0 / r)
\end{eqnarray}
where  $h_0 = 285$~nm, is the distance between the CNT and the back-gate, $r = 1-3$~nm the radius of a single-walled CNT, $V^{m}$ is the peak voltage at resonance and $V^{bg}$ is the background voltage.
Using the experimental parameters used for the measurements shown in Fig. 4 of the main text together with the ratio of the observed electrical and mechanical mixing signal amplitudes, we estimate an amplitude (peak to peak) of 0.14~nm for \textbf{a} and 0.7~nm for \textbf{b}.

\clearpage

\section*{Supplementary Note 3. Ringdown measurements and modelling the mechanical ringdown response.}
\label{rd:a:qrfit1}

To obtain the ringdown response, the following procedure was used: repeated pulses are applied to the switch (see circuit Fig.~1 of the main text) to turn the drive signal on and off using an RF switch.
At the same time the quadratures from the lock-in amplifier are measured by an oscilloscope.
The oscilloscope is triggered by the same pulse which drives the switch.
We also average the measured quadrature components typically 10000 times, to improve the signal to noise ratio.

Supplementary Fig. \ref{rd:meas1} shows the result for the rotated $X$ quadrature for two cases:
the green dashed curve shows the response for which the resonator was driven off-resonance ($f$ = 301.4~MHz), whereas the solid blue curve shows the response for which the resonator was actuated on-resonance ($f$ = 301.596~MHz). Both curves have been offset in the y-direction such that they approach zero with time.
The off-resonance curve $f$ = 301.4~MHz shows the lock-in related ringdown of the electrical background, independent of the mechanical resonance.\\

\textbf{Modelling the mechanical ringdown response.}
\label{rd:a:qrfit2}
To obtain an accurate ringdown \Qr~factor, we fit the measured ringdown response to a modelled ringdown response.
Electrical offsets can be quantified and subtracted from the rotated lock-in quadrature, prior to fitting.
In the modelled ringdown response, we include the effects of dephasing, as fluctuations in the resonance frequency ($\omega_0$) with time.  
We assume that these fluctuations occur on a time-scale slower than the the mechanical response time of the resonator ($2 \pi  \frac{Q_{\text{R}}}{\omega_0}$), but faster that the total averaging time of the measurement (typically 5 seconds).
Additionally, as the measured ringdown response is convoluted by the filter response function of the lock-in, we convolve the simulated data with the lock-in filter response function.
The lock-in response function is acquired by sending in a step-function signal into the lock-in, and its output is recorded by an oscilloscope.
The lock-in filter function is equal to the derivative of the measured output signal shown in Supplementary Fig. \ref{rd:lock-inresponse}.
Convolving the modelled data with the lock-in filter function allows us to fit the simulated data to the measured data in order to independently extract the ringdown time.\\

\textbf{Modelling of quadrature data.}
We consider the signals for the $X$ and $Y$ quadrature components from the lock-in,
originating from the mechanical response of the resonator without any noise and before getting averaged by the oscilloscope.
For now, we ignore effects due to the lock-in time constant.
For a driven response, the real and imaginary component of Eq. \ref{rd:eqHHO} are proportional to the $X$ and $Y$ quadrature magnitudes (which are measured by the oscilloscope).

From the measured data, we can fit the spectral \Qs~and resonance frequency of the resonator using Eq. \ref{rd:eqHHOabs}.
By inserting the fitted results for the spectral \Qs~and resonance frequency $\omega_0$ 
into Eq. \ref{rd:eqHHO2}, we can calculate the real ($X_d(\omega_d)$) an imaginary ($Y_d(\omega_d)$) results as a function of drive frequency.
These results for $X_d(\omega_d)$ and $Y_d(\omega_d)$ are proportional to these $X$ and $Y$ quadratures, for times when the driving is activated:
\begin{eqnarray}
H_{HO}(\omega) = k_R \frac{u(\omega)}{F(\omega)} = \frac{\omega_0^2}{\omega_0^2-\omega^2 + i \omega \omega_0 /Q} \label{rd:eqHHO2}
\end{eqnarray}
\begin{eqnarray}
X_{Ld}^{mech} &\propto & X_d(\omega_d)
\label{rd:eq:xmodel1}\\
Y_{Ld}^{mech} &\propto & Y_d(\omega_d)
\label{rd:eq:ymodel1}
\end{eqnarray}

For $t > 0$, when driving is switched off, two important effects happen to the mechanical signal.
First, the amplitude decays with time given by the ringdown \Qr~factor.
Second, for off resonance driving frequencies, the resonators frequency will be $\omega_0$ rather than $\omega_d$, which results in an oscillation of the $X$ and $Y$ quadrature signals.
By using the same method as before (Supplementary Note 1.), together with Eqs. \ref{rd:eq:xmech3} and \ref{rd:eq:ymech3} we get:
\begin{eqnarray}
X_{Lr}^{mech} \propto & 
\left(  
X_d(\omega_d) \cos(\omega_{beat} t) +
Y_d(\omega_d) \frac{\omega}{\omega_0} \sin(\omega_{beat} t)
\right)
e^{\frac{-\omega t}{2Q_{\text{R}}}}
\label{rd:eq:xmodel2}
\\
Y_{Lr}^{mech} \propto & 
\left(  
X_d(\omega_d) \sin(\omega_{beat} t) + 
Y_d(\omega_d) \frac{\omega}{\omega_0} \cos(\omega_{beat} t)
\right)
e^{\frac{-\omega t}{2Q_{\text{R}}}}
\label{rd:eq:ymodel2}
\end{eqnarray}
, where $e^{\frac{-\omega t}{2Q_{\text{R}}}} = e^{-\frac{t}{2 \tau}}$.
By using these four equations, we can obtain a numerical result of an amplitude which is proportional to the $X$ and $Y$ quadrature output of the lock-in (assuming the lock-in time-constant is equal to zero).\\

\textbf{Modelling of dephasing and averaged quadrature data.}
From the spectral fit of the measured data, we obtain the resonance frequency and spectral \Qs~factor. 
Using this resonance frequency and a given ringdown \Qr~factor, we can calculate a quadrature response as a function of drive frequency and time.
Without added dephasing the spectral \Qs~factor is equal to \Qr~factor.
We assume that the spectral decay rate $Q_S^{-1}$ can be written as the sum of the intrinsic dissipation 
rate $Q_\text{R}^{-1}$ and a dephasing rate $Q_\text{D}^{-1}$ i.e. $Q_\text{S}^{-1} = Q_\text{R}^{-1} + Q_\text{D}^{-1} $.
Dephasing is added by convolving this map with an amplitude distribution along the frequency axis which corresponds to a \Qd~factor, is given by:
\begin{eqnarray}
Q_\text{D} =  \frac{1}{Q_\text{S}^{-1} - Q_\text{R}^{-1}},
\end{eqnarray}
The result is a response map whose spectral \Qs~matches that of the measured data and \Qr~is a free variable.
The measured ringdown data is convolved by the lock-in filter function.
Once the response map is also convolved with the lock-in response filter function \Qr~can be fitted for.\\

\textbf{Extracting and convolving with the lock-in response filter function.}
To complete the simulated data set we add the effects due to the lock-in time constant.
The time constant defines the shape of the filter function of the lock-in in the time domain. 
To obtain the filter function, we generate a step function signal which is mixed by a reference signal from the lock-in amplifier. 
This signal is then fed into the input of the lock-in and the two quadratures $X$ and $Y$ from the lock-in output are recorded by the oscilloscope, alongside the step function (Supplementary Fig. \ref{rd:lock-inresponse}).
By taking the derivative of the output step function recorded by the oscilloscope, we obtain a lock-in filter function.
To verify that this filter function is correct, we convolve the recorded step function and see if the shape of the signal matches that of the measured signal.
Once the filter function is obtained, we convolve it with the simulated data set along the time axis.
Supplementary Fig. \ref{rd:simdataset} shows such the simulated and measured dataset.\\

\textbf{Fitting \Qr~using the modelled data.}
To fit the ringdown \Qr~factor, the previous steps have been repeated for different ringdown quality factors, while comparing the time domain at the resonance positions.
By minimising the mean square deviation ($\chi ^2$) between the measured and the simulated data set, we can fit for \Qr.
Supplementary Fig. \ref{rd:simdataset2} shows such a plot, which was recorded to obtain a rough estimate in fitting error.
A python code performing this, is available on Github here: \\
\url{https://github.com/benschneider/dephasing_ringdown_sim}\\

\textbf{Gate dependent measurements.}
Supplementary Fig. \ref{rd:A} shows the electronic and mechanical DC response of the region.

\section*{Supplementary Methods}
Fabrication starts with the sputtering of 50~nm tungsten onto the whole substrate.
To create the local gates, this layer is etched using an SF$_6$/He plasma, using a 300~nm thick NEB-22 resist mask.
The local gates are then covered under a 200~nm plasma enhanced chemical vapour deposited silicon dioxide layer. Subsequently, 70~nm of an 60-40 molybdenum rhenium alloy is deposited by sputtering.
Source and drain electrodes are defined by applying a PMMA/W/S1813 resist mask and subsequently etching 
with an SF$_6$/He, then O$_2$ and  finally an SF$_6$/He plasma.
Catalyst islands are then patterned on the substrate and nanotubes are grown using a CVD method at 900\degree C \cite{schneider2012coupling}.

\clearpage
